\documentclass{elsart}

\usepackage{graphicx}
\usepackage{epsfig}

\usepackage{amssymb}
\usepackage{eucal}

\newcommand{\gev}{$\mathrm{(GeV/c)}^{2}$}
\newcommand{\qsq}{$\mathrm{Q^2}$}
\newcommand{\qvec}{$\mathrm{\vec{q}}$}
\newcommand{\he}{$\mathrm{^3He}$}
\newcommand{\vhe}{$\mathrm{^3\overrightarrow{\mathrm{He}}}$}
\newcommand{\hee}{$\mathrm{^3\overrightarrow{\mathrm{He}}(\vec{e},e')}$}
\newcommand{\heen}{$\mathrm{^3\overrightarrow{\mathrm{He}}(\vec{e},e'n)}$}
\newcommand{\heep}{$\mathrm{^3\overrightarrow{\mathrm{He}}(\vec{e},e'p)}$}
\newcommand{\gen}{$G_{en}$}

\newcommand{\ape}{$A_{\perp}$}
\newcommand{\apa}{$A_{\parallel}$}
\newcommand{\et}{{\em et al.}}

\begin{document}

\begin{frontmatter}
%
\title{Final State Interaction Effects in 
 \mbox{$\mathrm{^3\vec{He}(\vec{e},e'p)}$}}


\author[a]{C.~Carasco}
\author[g]{J.~Bermuth}
\author[f]{P.~Merle}
\author[f]{P.~Bartsch}
\author[f]{D.~Baumann}
\author[f]{R.~B{\"o}hm}
\author[k]{D.~Bosnar}
\author[f]{M.~Ding}
\author[f]{M.O.~Distler}
\author[f]{J.~Friedrich}
\author[f]{J.M.~Friedrich}
\author[d]{J.~Golak}
\author[b]{W.~Gl{\"o}ckle}
\author[a]{M.~Hauger}
\author[g]{W.~Heil} 
\author[f]{P.~Jennewein}
\author[a]{J.~Jourdan\corauthref{cor1}} 
\corauth[cor1]{Corresponding author, e-mail: Juerg.Jourdan@unibas.ch}
\author[i]{H.~Kamada}
\author[h]{A.~Klein}
\author[c]{M.~Kohl}
\author[f]{K.W.~Krygier} 
\author[f]{H.~Merkel}
\author[f]{U.~M\"uller}
\author[f]{R.~Neuhausen}
\author[j]{A.~Nogga}
\author[a]{Ch.~Normand}
\author[g]{E.~Otten}
\author[f]{Th.~Pospischil}
\author[e]{M.~Potokar}
\author[a]{D.~Rohe}
\author[f]{H.~Schmieden}
\author[g]{J.~Schmiedeskamp}
\author[f]{M.~Seimetz}
\author[a]{I.~Sick}
\author[e]{S.~\v{S}irca}
\author[d]{R.~Skibi{\'n}ski}
\author[a]{G.~Testa}
\author[f]{Th.~Walcher}
\author[a]{G.~Warren}
\author[f]{M.~Weis}
\author[d]{H.~Wita{\l}a}
\author[a]{H.~W{\"o}hrle}
\author[a]{M.~Zeier}

\address[a]{Dept. f\"ur Physik und Astronomie, Universit\"at Basel, Switzerland}
\address[b]{Insitut f\"ur Theoretische Physik II, Ruhr--Universit\"at Bochum, Germany}
\address[c]{Institut f\"ur Kernphysik, Technische Universit\"at Darmstadt, Germany}
\address[d]{Institute of Physics, Jagiellonian University, Krak{\'o}w, Poland}
\address[e]{Institute Jo\v{z}ef Stefan, University of Ljubljana, Ljubljana, Slovenia}
\address[f]{Insitut f\"ur Kernphysik, Johannes Gutenberg--Universit\"at, Mainz, Germany}
\address[g]{Insitut f\"ur Physik, Johannes Gutenberg--Universit\"at, Mainz, Germany}
\address[h]{Dept. of Physics, Old Dominian University, Norfolk, USA}
\address[i]{Dept. of Physics, Kyushu Institute of Technology, 
Tobata, Kitakyushu, Japan}
\address[j]{Dept. of Physics, University of Arizona, Tucson, Arizona, USA}
\address[k]{Dept. of Physics, University of Zagreb, Croatia} 

\begin{abstract}
Asymmetries in quasi--elastic \heep\ have been measured at a momentum 
transfer of 0.67~\gev\ and are compared to a calculation which takes 
into account relativistic kinematics in the final state and a 
relativistic one-body current operator. With an exact solution of the Faddeev 
equation for the $\mathrm{^3He}$--ground  state and an approximate treatment 
of final state interactions in the continuum good agreement is 
found with the experimental data. 
\end{abstract}

\vspace*{-4mm}

\end{frontmatter}


{\bf Introduction:} \hspace*{0.01cm}
\label{intro}
Investigations of the structure of the nuclear three-body system have recently
attracted much interest. Modern three-body calculations allow for a quantitative 
description of this system not only of the ground state
but also for the continuum states. Results of such calculations open the 
possibility to test our understanding of the three-body system, the role 
of three-body forces, and non--nucleonic degrees of freedom by using 
continuum observables, quantities that obviously 
have a much richer structure and contain additional information. 
These calculations have reached a high degree of sophistication, and several 
''exact'' calculations are available today \cite{Gloeckle96,Witala01}. 

Recently, \he\ became also important for studies of nucleon form factors. 
Due to the lack of free neutron targets, only 
neutrons bound in light nuclei can be studied. The main advantage of \he\
lies in the fact that for the major part of the ground state wave
function the spins of the two protons are antiparallel so that spin
dependent observables are dominated by the neutron \cite{Golak02b}.

When using \he\ as a neutron target, nuclear structure effects such as final state 
interactions (FSI), meson exchange currents (MEC), and relativistic effects 
must be carefully considered \cite{Schulze93}. 
With the calculations available today such corrections can 
be performed quantitatively at low \qsq\ as was demonstrated in the recent 
electromagnetic form factor experiments by Becker \et\ 
and Xu \et\ \cite{Becker99,Xu00}.
However, given that the calculations were performed in a non-relativistic 
framework, a ``rigorous'' treatment of these corrections at higher \qsq\ was 
not at hand. This represents a serious difficulty for experiments aiming
at the electric neutron form factor, \gen.

The present paper reports about a new, less rigorous approach to correct for
nuclear effects at high \qsq. The theoretical results are 
compared with asymmetries measured in quasi-elastic \heep-scattering at 
\qsq=0.67\gev.

{\bf Theory:} \hspace*{0.01cm}
\label{theory}
The calculation is based on an exact, but non--relativistic \he\ ground state
wave function. To obtain the
matrix elements relativistic kinematics and a relativistic single nucleon 
current operator are used. Thereby the final state includes re--scattering 
terms to first order in the nucleon-nucleon (NN) t--matrix. 
Results for the AV18 NN--potential \cite{Wiringa95} will be presented. The dependence
on the NN--interaction is studied with a calculation which employs the CD--Bonn 
NN--potential \cite{Machleidt96}. In order to provide insight into the importance
of relativity additional calculations with a non--relativistic 
current or with non--relativistic kinematics are performed. All calculations use the
parameterization by H\"ohler to describe the electromagnetic form factors of
the nucleons \cite{Hoehler76}.

\begin{figure}[htb]\centering
\includegraphics[scale=0.75,clip]{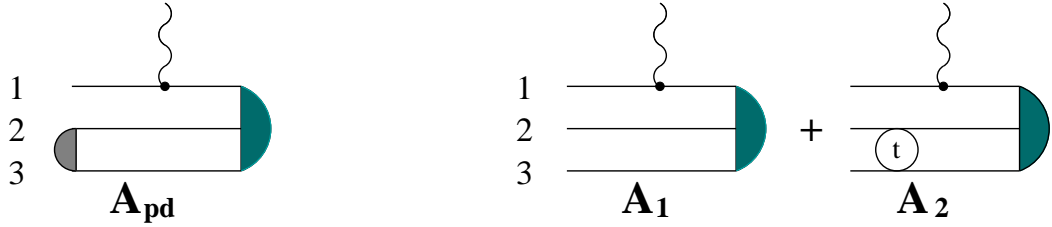}
\caption{\label{fsi_fig}
Diagrammatical representation of the two-body (left)
and three-body (right) breakup of $^3$He. 
The curly lines denote the photon coupling to nucleon 1.
The large(small) semi-circles depict the $^3$He(deuteron) bound state.
In $A_1$ FSI is neglected, in $A_2$ 
the scattering operator $t$ acts only in the subsystem $(23)$.
Note, that for the two-body breakup there is no diagram corresponding to $A_2$.
}
\end{figure}

As mentioned above, $\mathrm{^3\overrightarrow{\mathrm{He}}(\vec{e},e'N)}$
at large \qsq\ does not allow for a rigorous treatment 
of FSI based on the Faddeev-like integral equation \cite{Gloeckle96}.
When the center-of-mass energy of the three-nucleon (3N) system is well above the pion 
production threshold the usual potential approach is not valid. However, in 
quasi--elastic kinematics the focus is mostly on the region of phase space, where one of the 
nucleons is struck with a high energy and momentum and leaves the remaining 
two-nucleon system with a rather small internal energy. Thus, one may hope that 
approximations shown in figure \ref{fsi_fig} will be justified. 

Let us first consider the three-body breakup of $^3$He.
The amplitude $A_1$ takes a very simple form 
\begin{equation}
A_1 = \langle {\vec p}_1 m_1 \nu_1 {\vec p}_2 m_2 \nu_2 {\vec p}_3 m_3 \nu_3
      \mid
       j ( {\vec Q} , 1) \mid \Psi_b {\vec P} M M_T \rangle ,
\end{equation}
where $ {\vec p}_i$ are the individual nucleon momenta and ${m_i}$ (${\nu_i}$) 
their spin (isospin) projections. ${\vec P}$ is the total momentum and
$M$ ($M_T$) the spin (isospin) magnetic 
quantum number of the initial $^3$He bound state. The single nucleon current
$j ( {\vec Q} , 1)$ acts only on nucleon 1. 
Choosing the laboratory frame ${\vec P}=0$ and the standard representation 
of the 3N bound state in the basis of relative Jacobi momenta $\vec p$ and $\vec q$ 
one gets 
\begin{eqnarray}
A_1 = & \delta ( {\vec p}_1 + {\vec p}_2 + {\vec p}_3 - {\vec Q} ) 
\sum_{{m_1}'} j( {\vec p}_1 , {\vec p}_1 - {\vec Q} ; m_1 , {m_1}'; \nu_1)
\nonumber \\
& \langle {\vec p} {\vec q} {{m_1}'} m_2 m_3 \nu_1 \nu_2 \nu_3
      \mid \Psi_b M M_T \rangle ,
\end{eqnarray}
where $ {\vec p} = \frac12 ( {\vec p}_2 - {\vec p}_3 ) $, 
      $ {\vec q} = {\vec p}_1 - {\vec Q} $.
Finally we use the partial wave decomposition of the bound state 
in the basis $ \mid p q \alpha \rangle$ (see \cite{Gloeckle87}) and arrive at
\begin{eqnarray}
A_1 = \delta ( {\vec p}_1 + {\vec p}_2 + {\vec p}_3 - {\vec Q} ) \
      \delta_{\nu_1 + \nu_2 + \nu_3 , M_T } \ 
\sum_{{m_1}'} j( {\vec p}_1 , {\vec p}_1 - {\vec Q} ; m_1 , {m_1}'; \nu_1) 
          \nonumber \\
\sum_{{\alpha}'} \ \sum_{{\mu}'} \ 
     C(j',I',\frac12; {\mu}',M-{\mu}',M) \
     C(l',s',j';{\mu}'-m_2-m_3,m_2+m_3,{\mu}') 
           \nonumber \\
     C(\frac12, \frac12, s'; m_2 , m_3, m_2+m_3) \
     C({\lambda}', \frac12, I' ; M-{\mu}'-{m_1}' , {m_1}' , M-{\mu}' ) 
            \nonumber \\
     C(t',\frac12, \frac12 ; \nu_2 + \nu_3, \nu_1 , \nu_1 + \nu_2 + \nu_3 ) \
     C(\frac12, \frac12, t'; \nu_2 , \nu_3, \nu_2+\nu_3) 
         \nonumber \\
  Y_{ l', {\mu}'-m_2-m_3 } ( {\hat p} ) \
      Y_{ {\lambda}', M-{\mu}'-{m_1}' } ( {\hat q} ) \
       \langle p q {\alpha}' \mid \Psi_b \rangle . \ \ \ \ \ \ \
\end{eqnarray}
 
The amplitude $A_2$ is given by
\begin{equation}
A_2 = \langle {\vec p}_1 m_1 \nu_1 {\vec p}_2 m_2 \nu_2 {\vec p}_3 m_3 \nu_3
      \mid t_{23} \, G_0 \,
       j ( {\vec Q} , 1) \mid \Psi_b {\vec P} M M_T \rangle
\end{equation}
and can be written as

\begin{eqnarray}
A_2 & = & \delta ( {\vec p}_1 + {\vec p}_2 + {\vec p}_3 - {\vec Q} ) \
      \delta_{\nu_1 + \nu_2 + \nu_3 , M_T } \
      \delta_{\nu_1 , {\nu_1}'} \
\sum_{{m_1}'} j( {\vec p}_1 , {\vec p}_1 - {\vec Q} ; m_1 , {m_1}'; \nu_1) 
          \nonumber \\
&& \int \, d {\vec p}^{\ '} \, \sum_{ {m_2}',{m_3}'} \, \sum_{ {\nu_2}',{\nu_3}'} \,
         \delta_{\nu_2 + \nu_3 , {\nu_2}' + {\nu_3}' } \
\langle {\vec p} m_2 m_3 \nu_2 \nu_3 \mid t (z) \mid  {\vec p}^{\ '} {m_2}' {m_3}' {\nu_2}' {\nu_3}' \rangle \nonumber \\
&& \frac1{E - E(p_1, p_{23}, p') + i \epsilon } \
\langle {\vec p}^{\ '} {\vec q} {{m_1}'} {m_2}' {m_3}' {\nu_1}' {\nu_2}' {\nu_3}'
      \mid \Psi_b M M_T \rangle ,
\end{eqnarray}
where $ {\vec q} = {\vec p}_1 - {\vec Q} $ 
and  $ {\vec p}_{23} = {\vec p}_2 + {\vec p}_3 $.
The total energy $E$ of the 3N system can be expressed as 
\begin{eqnarray}
E = \omega + m_{^3{\rm He}} 
  = \sqrt{ m_N^2 + p_1^2 } + \sqrt{ 4 ( m_N^2 + p^2 ) + p_{23}^2 } 
     \nonumber \\
  \approx 
  \sqrt{ m_N^2 + p_1^2 } + \sqrt{ 4 m_N^2 + p_{23}^2 } 
        + \frac{p^2}{ \sqrt { m_N^2 + \frac14 p_{23}^2 } } 
      \ \equiv E(p_1, p_{23}, p) .
\label{eq6}
\end{eqnarray}
with $m_N$ the nucleon mass. In Eq.~(\ref{eq6}) 
${\vec p}$ is the (relativistic) relative momentum between nucleons 2 and 3 
calculated in the frame where the total momentum of the $(23)$ pair is zero.
It agrees, however, to a good approximation with the standard (nonrelativistic) 
definition $ {\vec p} = \frac12 ( {\vec p}_2 - {\vec p}_3 ) $.
Consequently, the $(23)$ subsystem internal energy which enters 
in the nonrelativistic t-matrix calculation is taken as 
\begin{equation}
z = E -  \sqrt{ m_N^2 + p_1^2 } - \sqrt{ 4 m_N^2 + p_{23}^2 }
\end{equation}

In a final step both the bound state wave function and the t-matrix are given in the
partial wave basis, which yields 
\begin{eqnarray}
A_2 = \delta ( {\vec p}_1 + {\vec p}_2 + {\vec p}_3 - {\vec Q} ) \
      \delta_{\nu_1 + \nu_2 + \nu_3 , M_T } \
\sum_{{m_1}'} j( {\vec p}_1 , {\vec p}_1 - {\vec Q} ; m_1 , {m_1}'; \nu_1)
           \nonumber \\
\sum_{ l s j \mu t } \
   C(l,s,j;{\mu}-m_2-m_3,m_2+m_3,{\mu}) \ 
   C(\frac12, \frac12, s; m_2 , m_3, m_2+m_3) \
           \nonumber \\
  C(t,\frac12, \frac12 ; \nu_2 + \nu_3, \nu_1 , \nu_1 + \nu_2 + \nu_3 ) \
 C(\frac12, \frac12, t; \nu_2 , \nu_3, \nu_2+\nu_3) \ 
           \nonumber \\
Y_{ l, {\mu}-m_2-m_3 } ( {\hat p} ) \
\sum_{\bar{l}} \, \sum_{{\alpha}'} \, \delta_{l' \bar{l}} \,
               \delta_{s' s} \,
               \delta_{j' j} \,
               \delta_{t' t} \,
     C(j,I',\frac12; {\mu},M-{\mu},M) \nonumber \\
     C({\lambda}', \frac12, I' ; M-{\mu}-{m_1}' , {m_1}' , M-{\mu} ) \
 Y_{ {\lambda}', M-{\mu}-{m_1}' } ( {\hat q} ) 
           \nonumber \\
\int \, d p' \, {p'}^{\ 2} \ 
\frac
{ \langle p' q {\alpha}' \mid \Psi_b \rangle }
{E - E(p_1, p_{23}, p') + i \epsilon } \
\langle p (l s )j t \mid t (z) \mid p' (l' s')j t \rangle .
\end{eqnarray}

The amplitude for the two-body breakup of $^3$He, $A_{pd}$, is 
given as
\begin{equation}
A_{pd} = \langle {\vec p}_1 m_1 \nu_1 \, \phi_d {\vec p}_d m_d 
      \mid
       j ( {\vec Q} , 1) \mid \Psi_b {\vec P} M M_T \rangle ,
\end{equation}
where $\phi_d$ is the deuteron state with the spin magnetic quantum number $m_d$ 
and laboratory momentum ${\vec p}_d$. In the next step one gets 
\begin{eqnarray}
A_{pd} = & \delta ( {\vec p}_1 + {\vec p}_d - {\vec Q} ) \ \delta_{ \nu_1 M_T } \
\sum_{{m_1}'} j( {\vec p}_1 , {\vec p}_1 - {\vec Q} ; m_1 , {m_1}'; \nu_1) 
\nonumber \\
& \langle {\vec q} {{m_1}'} \nu_1 \phi_d {\vec p}_d m_d \mid \Psi_b M M_T \rangle ,
\ \ \ \ \ \ \ 
\end{eqnarray}
where $ {\vec q} = {\vec p}_1 - {\vec Q} $.
Partial wave expansion of the deuteron and $^3$He bound state 
leads to 
\begin{eqnarray}
A_{pd} & = & \delta ( {\vec p}_1 + {\vec p}_d - {\vec Q} ) \
      \delta_{\nu_1 M_T } \
\sum_{{m_1}'} j( {\vec p}_1 , {\vec p}_1 - {\vec Q} ; m_1 , {m_1}'; \nu_1)
           \nonumber \\
&& \sum_{{\alpha}'} \ 
( \delta_{l' 0} \, + \, \delta_{l' 2} ) \
               \delta_{s' 1} \,
               \delta_{j' 1} \,
               \delta_{t' 0} \,
     C(j',I',\frac12; m_d, M-m_d, M) \
       \nonumber \\
&&     C({\lambda}', \frac12, I' ; M-m_d-{m_1}' , {m_1}' , M-m_d ) \
   Y_{ {\lambda}', M-m_d-{m_1}' } ( {\hat q} ) 
       \nonumber \\
&& \int \, d p' \, {p'}^{\ 2} \ \varphi_{l'} (p') \,
     \langle p' q {\alpha}' \mid \Psi_b \rangle .
\end{eqnarray}

The single nucleon current matrix elements
$ j( {\vec p}_1 , {\vec p}_1^{\ '} ; m_1 , {m_1}'; \nu_1) $
($\nu_1$ decides whether the photon couples 
to the proton or to the deuteron) are taken completely relativistically, i.e.,
\begin{eqnarray}
j( {\vec p}_1 , {\vec p}_1^{\ '} ; m_1 , {m_1}')  \equiv 
j^\mu ( {\vec p}_1 , {\vec p}_1^{\ '} ; m_1 , {m_1}')
\ \ \ \ \ \ \ \ \ \ \ \ \ \ \ \ \ \ \ \ \ \ \ \ \ \ \ \ \ \ \ \ &&
 \\
 = 
                 \sqrt{ \frac{m_N}{ \sqrt{m_N^2 + p^2 }}} \
                 \sqrt{ \frac{m_N}{ \sqrt{m_N^2 + {p'}^{\ 2 }}}} 
{\bar u}(p m_1) \left( F_1 \gamma^\mu + i F_2 \sigma^{\mu \nu} (p - p')_\nu \right)
u(p' {m_1}') && \nonumber 
\end{eqnarray}
In the case of  $A_{pd}$ only the proton single nucleon current 
contributes ($ \nu_1 = M_T = \frac12$).
The amplitudes $A_1$, $A_2$ and $A_{pd}$ are used 
to calculate the response functions entering the cross sections
and the polarization observables. 

\indent
To simplify integrations over the
unobserved parameters of the final 3N system we change the variables
according to \cite{Gloeckle86}:
\begin{equation}
d^3 p_1 \, d^3 p_2 \, d^3 p_3 \ = \ 
{\mathcal{I}} \, d^3 p_1 \, d^3 p_{23} \, d^3 p ,
\end{equation}
with
${\mathcal{I}} = {\frac{ 4 E_2 E_3 }{ ( E_2 + E_3 ) \sqrt{ (E_2 + E_3)^2 - p_{23}^2 } }}$
($E_i$ is the total energy of the i$^{th}$ nucleon).

{\bf Experiment:} \hspace*{0.01cm}
\label{exp}
Asymmetries of the \heep--reaction have been measured in quasi--elastic 
kinematics at a four momentum transfer of \qsq=0.67\gev. As the spin
asymmetry of protons in \he\ is very small \cite{Schulze93} one can
expect that the asymmetries of the \heep--reaction are very sensitive to 
FSI and provide a significant test of the calculations. 

For an arbitrary direction of the target spin the asymmetry is given by
\begin{equation}
A(\theta^*,\phi^*) = \frac{1}{P_e \cdot P_t} \cdot \frac{N^+ - N^-}{N^+ + N^-}
\end{equation}
with $\theta^*,\phi^*$ the polar and azimuthal angle of the target spin 
direction with respect to the three momentum transfer \qvec. The polarizations
of beam and target are given by $P_e$ and $P_t$ and the 
normalized \heep\ events for positive (negative) electron helicity are 
$N^+ (N^-)$. The two asymmetries \apa\ and \ape\ measured in 
the present work are $A_{\perp} = A(90^o,0^o)$ and $A_{\parallel} = A(0^o,0^o)$
\cite{Raskin89}.

The setup of the experiment was very similar to the one described by 
Rohe \et\ \cite{Rohe99}. A polarized continuous wave electron beam with an energy 
of 854.5~MeV was incident on a glass cell filled with 
polarized \he. Longitudinally polarized electrons of $\sim$80\% polarization were 
produced with a strained layer GaAsP crystal at a typical current of $10~\mu A$
\cite{Aulenbacher97}. Spectrometer A with a solid angle of 28~msr and a 
momentum acceptance of 20\% \cite{Blomqvist97} was used to detect the scattered 
electron at a scattering angle of 78.6~deg. The struck protons were detected in 
coincidence with an array of plastic scintillator bars placed at 32.2~deg, 
the direction of \qvec\ for an energy transfer $\omega$ of 368~MeV. 

The \vhe--target consisted of a spherical glass container with two cylindrical
extensions. The beam enters and exits through 25$\mu$m Cu--windows.
The cylindrical extensions allowed for an effective shielding of the background 
from the Cu--windows by positioning the windows outside of the acceptance 
of the spectrometer. 
The $^3{\rm He}$ gas was polarized by metastable optical pumping at pressures 
around 1~mbar and subsequently compressed by a two-stage titanium piston 
compressor  to 4~bar. The target polarization achieved was 
approximately 50\% \cite{Surkau97}. 

The entire target was enclosed in a rectangular box of 2~mm thick $\mu$-metal 
and iron. The box served as an effective shield for the 
stray field of the  magnetic  spectrometers and provided a 
homogeneous magnetic guiding field of $\approx 4 \cdot 10^{-4}$~T produced by three 
independent pairs of coils.
With additional correction coils a relative field gradient
of less than $5 \cdot 10^{-4}$\,cm$^{-1}$ was achieved. 
The setup allowed for an independent rotation of the target spin
in any desired direction by remote control. In order to reduce 
systematic errors, the spin of the target 
was circularly rotated in the scattering plane 
by 90$^o$  with respect to the direction of \qvec\ at regular intervals,
alternatively accumulating data for \apa\ and \ape.  

The hadron detector consists of an array of four layers of five 
vertically placed plastic scintillator bars with dimensions $50\times10\times10$~cm$^3$
preceded by two layers of 1~cm thick $\Delta$E detectors for particle identification. 
Every plastic bar was equipped with two Photo Multipliers (PM) on top and bottom. 
The detector was placed at 32.2$^o$ at a distance  of 160~cm from the 
target which resulted in a solid angle of 100~msr. The entire detector
was shielded with 10~cm Pb except for an opening towards the
target where the Pb--shield was reduced to 2~cm. 

As in the experiment by Rohe \et\ \cite{Rohe99} the product of target and beam
polarization was monitored during the data taking via determination of
the asymmetry for elastic \vhe$(\vec{e},e)$--scattering. The \he--form factors
are accurately known \cite{Amroun94} and the comparison of the calculated and
measured asymmetry allows for a precise determination of the polarization product $P_e \cdot P _t$.
The elastically scattered electrons were detected in spectrometer B at a scattering
angle of $\vartheta_e = 25^o$. This resulted in a polarization
product of $0.279\pm 0.010$ for runs with $A = A_{\parallel}$ or --\apa\
and $0.282\pm 0.003$ for $A = A_{\perp}$ or --\ape. The difference
of the error bar results from the different sensitivity of the measurement to the target spin 
direction. 

In addition, the time dependence of the polarization of the target cell was continuously 
measured during the experiment by Nuclear Magnetic Resonance, while the absolute 
polarization was measured by the method of Adiabatic Fast Passage \cite{Wilms97}.
The mean target polarization from these measurements was $0.356\pm 0.015$. 
From the elastic scattering data and the target polarization 
measurements a beam polarization of $P_e = 0.788\pm 0.036$ was determined which agreed
well with the result from a M{\o}ller polarimeter ($0.827\pm 0.017$).

{\bf Analysis:} \hspace*{0.01cm}
\label{anal}
In the off--line analysis, protons are defined as events with a hit in two
consecutive $\Delta$E detectors from the two $\Delta$E--detector planes. 
For the kinematics of the experiment the proton energies 
range from $\sim$280~MeV to $\sim$400~MeV. For this energy range protons reach 
at least the third bar layer. 
A correct mean time of at least three plastic bars is also required. 
For the combination of these cuts negligible background survives in the 
\heep\ coincidence time peak. 
The segmentation of the hadron detector and the up--down PM readout 
allow for the determination of the direction of the protons. The resolution
is $\sim$0.8$^o$ in both vertical and horizontal direction.

\begin{figure}[hbt]
\begin{center}
\includegraphics[scale=0.60,clip]{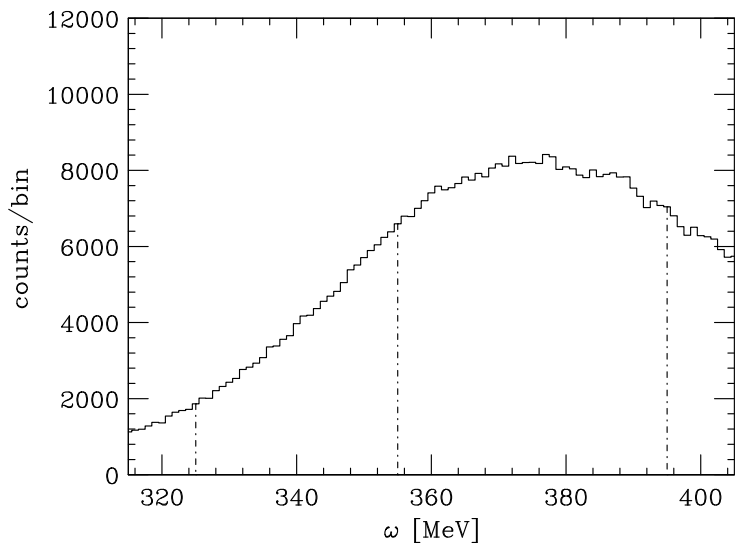}
\parbox{13cm}{\caption[]{
Experimental spectrum of the \heep--data as a function of the energy transfer 
$\omega$. The dot--dash lines indicate the two regions selected for the analysis.
}\label{ep_xsec}} 
\end{center} 
\end{figure}
In order to study the effect of FSI on the asymmetries in different kinematic
regions, the quasi--elastic peak is divided in two regions of $\omega$. 
One region covers the peak and 
therefore emphasizes low nucleon momenta whereas the other region
covers the low $\omega$ tail sensitive preferentially to high nucleon momenta. 
Figure \ref{ep_xsec} shows the $\omega$ spectrum of \heep--events and indicates 
the two kinematic regions. 
The events in each of the two regions are summed over the entire acceptance
of the out--of--plane angle of electron and proton and over the electron
scattering angle in a range from 75.8$^o$ to 81.8$^o$. 

\begin{figure}[hbt]
\begin{center}
\includegraphics[scale=0.595,clip]{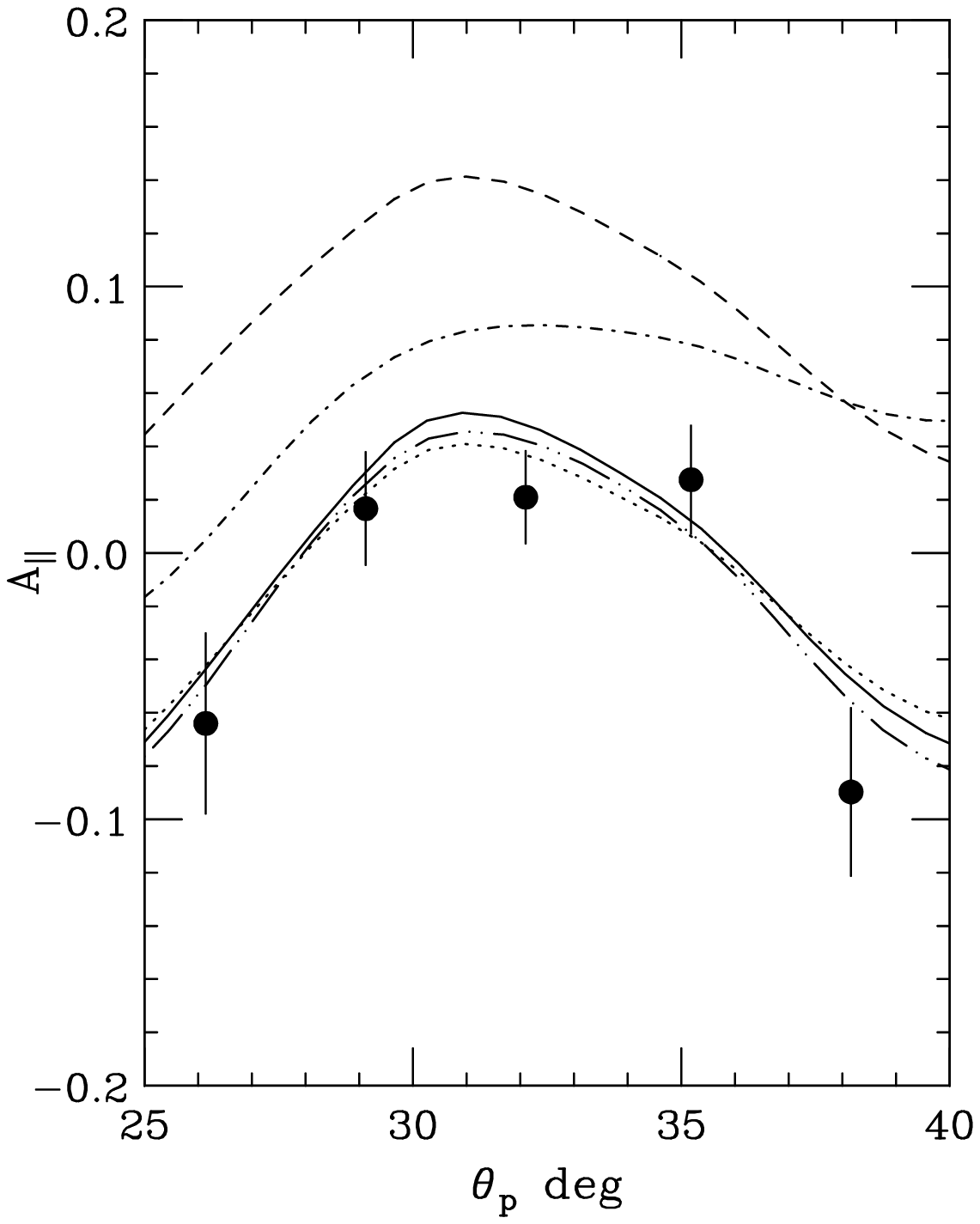}\includegraphics[scale=0.595,clip]{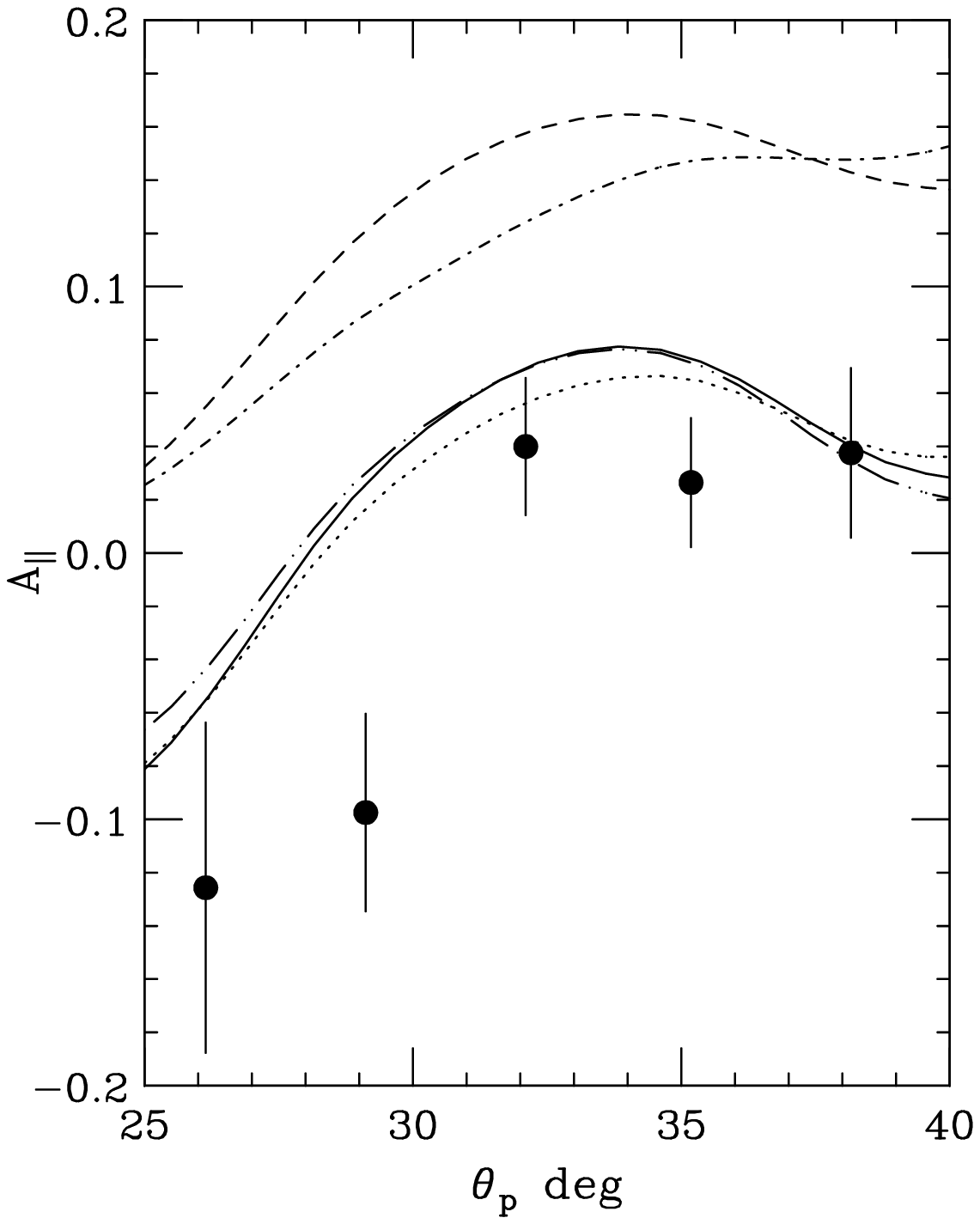}
\parbox{13cm}{\caption[]{
Experimental results of \apa\ for the quasi--elastic peak region (left) 
and the low $\omega$ tail region (right) as a function of the scattering angle of the knocked
out proton. The results of the full(PWIA) calculation are shown with solid(dashed) lines. The result of
the full calculation with a non--relativistic current (dot), the effect of a (v/c)$^2$ correction (dot-dot-dash) 
and the same with non--relativistic kinematics (dot-dash) are also shown.
}\label{eep_apa}} 
\end{center} 
\end{figure}

\begin{figure}[hbt]
\begin{center}
\includegraphics[scale=0.595,clip]{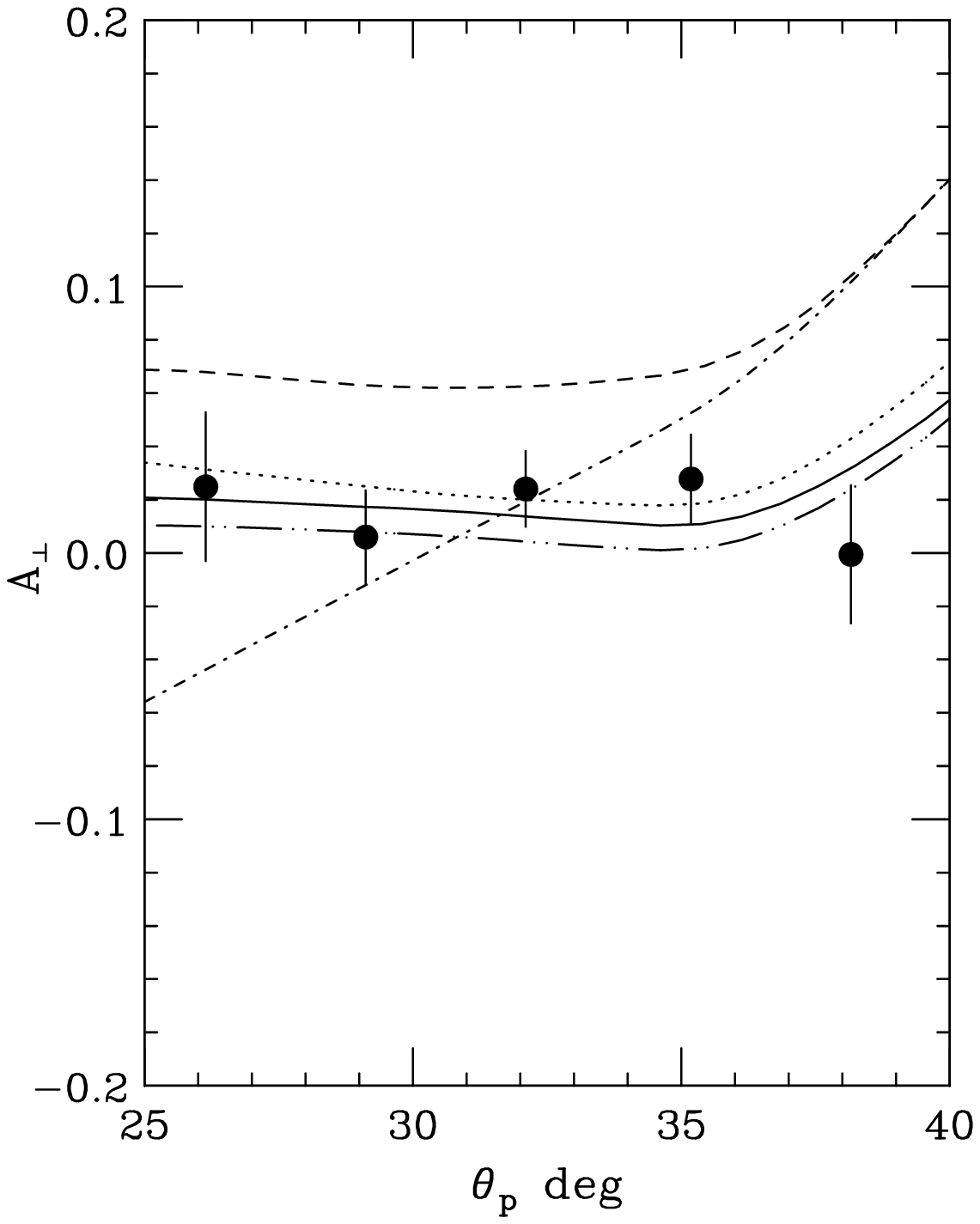}\includegraphics[scale=0.593,clip]{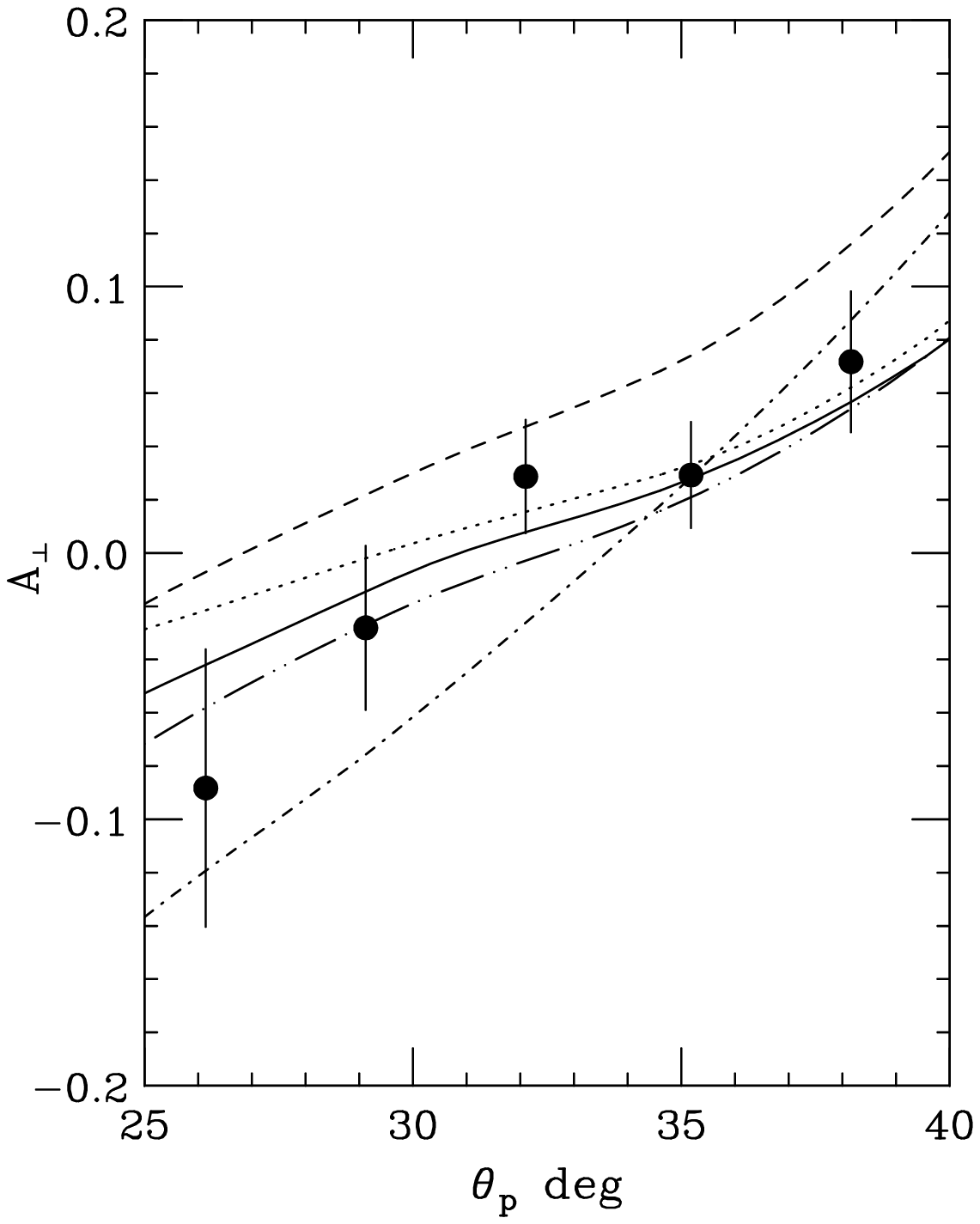}
\parbox{13cm}{\caption[]{
Same as figure \ref{eep_apa}, but for \ape.
}\label{eep_ape}} 
\end{center} 
\end{figure}

{\bf Results:} \hspace*{0.01cm}
\label{result}
The experimental results for \apa\ and \ape\ are shown in figure 
\ref{eep_apa} and \ref{eep_ape} 
compared to the results of the plane wave impulse approximation (PWIA) and the 
calculation including
the dominant FSI effect. As expected the asymmetries are small over the entire
kinematic region so that FSI effects appear prominently. 
Compared to the statistical accuracy (0.02--0.03) the 
systematic errors are small (0.01). The error of the deviation
of \qvec\ from its nominal direction, which is usually the dominant systematic error in
electric form factor determinations \cite{Rohe99}, is negligible here mainly 
because \apa\ and \ape\ are not very different from each other. 

The hadron detector does not allow for a 
high--resolution determination of the proton energy. In particular, it is not possible to 
distinguish two- and three-body breakup events. Accordingly,
the calculations have been integrated over the two-body
and over the first 26~MeV of the three-body breakup channel. 
For the highest accepted missing energy of 26~MeV, the cross section is smaller 
by at least one order of magnitude compared to the cross section at threshold. 
Extending the integration 
limit has no effect on the results. In addition, the calculated results are integrated over the 
experimentally accepted out--of--plane proton angle and the relevant $\omega$ 
range (figure \ref{ep_xsec}).

{\bf Conclusions:} \hspace*{0.01cm}
Figures \ref{eep_apa} and \ref{eep_ape} show that the PWIA calculations are in clear
disagreement with the experimental results. The same holds for the calculation which
does not account for relativistic kinematics. On the other hand, very good agreement 
between experiment and theory is found when including the $A_2$--term of figure \ref{fsi_fig}
and accounting for relativistic kinematics. 
This applies also for the results using the CD--Bonn NN--potential with negligible differences.
Small differences of the results are observed when the current is replaced by
a non--relativistic version or when a relativistic $(v/c)^2$--correction is added. 
The results indicate that at high \qsq, where complete non--relativistic
calculations are not applicable anymore, a good description of the data can be achieved 
taking into account relativistic kinematics and an approximate treatment of FSI--effects. 
The use of a relativistic current operator is less relevant. The result is important for experiments 
aimed at extracting fundamental properties of the neutron from asymmetry measurements of 
inclusive \hee\ or exclusive \heen\ reactions. The corrections for FSI--effects 
for these reactions can be reliably calculated within the approach presented here.

{\bf Acknowledgments:} \hspace*{0.01cm}
\label{ackn}
This work was supported by the Schweizerische Nationalfonds, Deutsche
Forschungsgemeinschaft (SFB 443), the Polish Committee for Scientific Research, 
the Foundation for Polish Science and
NSF grant \#PHY 0070858. The numerical calculations have
been performed on the Cray SV1 of the NIC, J\"ulich, Germany.



\end{document}